\documentclass[adp,fleqn]{w-art}
\usepackage{times}
\usepackage{w-thm}
\usepackage{graphicx}
\usepackage{amssymb}
\usepackage{amsbsy}
\usepackage{amsmath}
\usepackage{epsfig}
\usepackage{epstopdf}

\chardef\bslash=`\\ 

\def\be{\begin{equation}}
\def\ee{\end{equation}}
\def\bea{\begin{eqnarray}}
\def\eea{\end{eqnarray}}
\def\ba{\begin{array}}
\def\ea{\end{array}}

\begin{document}
\DOIsuffix{theDOIsuffix}
\Volume{12}
\Issue{1}
\Copyrightissue{01}
\Month{01}
\Year{2003}
\pagespan{1}{}
\Receiveddate{1 September 2009}
\Accepteddate{5 September 2009}


\title[Purely electronic transport and localization in the Bose glass]{Purely electronic transport and localization in the Bose glass}

\author[M.\ M\"uller]{M.\ M\"uller\footnote{Corresponding
     author \quad E-mail: {\sf markusm@ictp.it}, Phone: +39\,040\,2240\,350,
     Fax: +39\,040\,2240\,7350}\inst{1}}
\address[\inst{1}]{The Abdus Salam International Center for Theoretical Physics, Strada Costiera 11, 34014 Trieste, Italy}


\begin{abstract}
We discuss transport and localization properties on the insulating side of the disorder dominated superconductor-insulator transition, described in terms of the dirty boson model. Analyzing the spectral properties of the interacting bosons in the absence of phonons, we argue that the Bose glass phase admits three distinct regimes. For strongest disorder the boson system is a fully localized, perfect insulator at any temperature. At smaller disorder, only the low temperature phase exhibits perfect insulation while delocalization takes place above a finite temperature. We argue that a third phase must intervene between these perfect insulators and the superconductor. This conducting Bose glass phase is characterized by a mobility edge in the many body spectrum, located at finite energy above the ground state. In this insulating regime purely electronically activated transport occurs, with a conductivity following an Arrhenius law at asymptotically low temperatures, while a tendency to superactivation is predicted at higher $T$. These predictions are in good agreement with recent transport experiments in highly disordered films of superconducting materials.
\end{abstract}
\maketitle

\section{Introduction}
\label{intro}
The disordered superconductor-to-insulator transition (SIT)~\cite{Hebard,Goldman}
has recently attracted a lot of attention due to a series of surprising and puzzling transport experiments, especially on the insulating side of the quantum phase transition. In {\em strongly disordered} samples with nearly critical disorder, a large magnetoresistance peak is observed as the sample is tuned further away from the superconducting phase by a magnetic field~\cite{Shahar}. The latter~\cite{Hebard,Kapitulnik} and several further experimental indications in transport~\cite{Valles,TanParendo} and tunneling~\cite{Sacepe} are commonly interpreted as an indication for the survival of an electronic pairing gap close enough to the transition, despite the loss of global phase coherence in this kind of samples. This scenario is supported by numerical studies which suggest that the pairing gap and the existence of bosons (pairs) can survive both the disorder driven~\cite{Nandini} and the field-driven SIT~\cite{Dubi} (in the absence of Coulomb interactions). However, apart from the above described type of SI transitions, there is another class of systems with weaker disorder~\cite{Goldman,Kapitulnik} in which superconductivity is predominantly suppressed by Coulomb interactions and the concomitant destruction of electron pairing.

The experimental observations in strongly disordered samples motivate the assumption that the low energy physics in the vicinity of the SIT can be described by interacting bosons subject to a disorder potential. This forms a cornerstone of the dirty boson model for the SIT~\cite{Fisher},
which predicts that disorder (or a magnetic field) drive the superconducting state into an insulating, but compressible Bose glass phase. In sufficiently pure samples with a regular granular structure and commensurate filling a transition to a Mott insulating phase with a hard gap for charged excitations is predicted to exist at smaller hopping strength. However, such an incompressible phase has little reason to occur in strongly disordered thin films, independent of the presence or absence of self-generated granularity.

In this paper we adopt a purely bosonic description of the phase transition for dimensions $d\geq 2$. We focus on the properties of intrinsically electronic transport in the insulating Bose glass phase, where we strictly neglect electron-phonon coupling. The latter is empirically known to be rather weak in the relevant experimental systems~\cite{overheating}. The behavior of the intrinsic conductivity turns out to be surprisingly rich. Our analysis leads to a generic picture of the physics close to the SIT which we believe to be key to understanding several unexplained features of transport in Bose glasses. On one hand, it has been a long standing puzzle that strongly disordered superconducting films exhibit a simply activated (Arrhenius) resistance at low temperatures~\cite{activated} instead of variable range hopping, which is normally expected in disordered insulators.
 While the absence of variable range hopping may be blamed on the weak electron-phon coupling, most simple explanations for an Arrhenius resistance must be ruled out: The presence of a hard gap for bosons~\cite{Efetov} is excluded in the strongly disordered, compressible Bose glass. Further, an explanation based on nearest neighbor hopping is inconsistent with the observed magnitude of the pre-exponential factor of the conductivity.
 Finally, interpreting the activation energy as the energy to form unpaired electrons~\cite{FeigelmanIoffeKravtsov} is ruled out by the observed positive magnetoresistance. 

 Another interesting transport feature was discovered very recently. At ultralow temperatures, Bose glasses such as InO$_x$~\cite{Shahar} or TiN~\cite{Baturina} exhibit huge resistance jumps when driven beyond a certain threshold in voltage~\cite{overheating}. This phenomenon was explained as a thermal run-away into an overheated state of the electronic subsystem which is nearly decoupled from the phonons~\cite{overheatingtheory}. This interpretation relies crucially on the assumption that the resistance is sensitive to the electron temperature rather than the lattice temperature. As stressed in Ref.~\cite{overheatingtheory}, this is at odds with the standard mechanism of phonon assisted transport in insulators and implies that the transport in the Bose glass phase must essentially be activated by the electrons themselves. However, no microscopic mechanism was proposed.

In this paper, we argue that purely electronic transport and its simply activated form are both natural to expect close enough to the SIT for $d\geq 2$, provided that electron-phonon coupling plays a negligible role.

\section{The Bose glass}
\label{sect2}

For the general analysis of the phase diagram below, a specific microscopic picture of disordered bosons is not really essential. Nevertheless, in order to fix ideas it is helpful to argue in terms of the model considered by Ma and Lee~\cite{MaLee}. Let $\psi_i$ be the Anderson localized states of non-interacting electrons, each with a localization length of typical size $\xi$. We then restrict the Hilbert space of the full Hamiltonian (with an attractive BCS interaction) to configurations with all states $i$ either empty or doubly occupied with a spin singlet. This reduces the problem to a system of disordered hard core or pseudo-spin-$1/2$ bosons~\cite{AndersonPseudospins}:
\bea
\label{Ham}
H=-\sum_i(\epsilon_i-\mu) s_i^z +\sum_{i,j} (t_{ij} s_i^+ s_j^- +{\rm h.c.}).
\eea
Each state $i$ has a random energy $\epsilon_i$, the average density of states being $\nu$. The level spacing in the localization volume of bosons is thus $\delta_\xi=1/\xi^d\nu$, and $\mu$ denotes the chemical potential. In the Hamiltonian (\ref{Ham}) we have omitted a density-density interaction $\sum_{ i,j} J_{ij} s_i^z s_j^z$, as it would arise, e.g., from Coulomb interactions. However, it would not affect our general considerations as long as $J_{ij}$ are short ranged and weak enough so as not to introduce frustration and glassy effects.

Note that the attraction-induced hopping of bosons, $t_{ij}$, is appreciable only between spatially overlapping states, $t_{ij}\sim \exp[-|r_i-r_j|/\xi]$, with the center of mass coordinates $r_{i,j}$ of the states. If the states $i,j$ are within the same localization volume, $t_{ij}$ will typically decrease as a power law with distance in energy $\epsilon_i-\epsilon_j$. The superconductivity in such a model, built from nearly critical wavefunctions $\psi_i$, has recently been studied in great detail~\cite{Feigelman2}.

Let $t$ be the typical magnitude of $t_{ij}$ corresponding to nearest neighbors in space and energy. The dimensionless parameter $g\equiv \delta_\xi/t$ characterizes the strength of disorder as compared to the kinetic energy.
If $g\ll 1$, the disorder plays very little role and the physics will be equivalent to that of an infinite-$U$ Hubbard model at incommensurate filling: The bosons delocalize and condense into a superconducting state with infinite conductivity below a critical temperature $T_c(g)$ which decreases with growing $g$, see Fig.~\ref{fig:PhaseDia}. However, at some critical $g_c=O(1)$ the disorder becomes dominant and localizes the bosons, transforming the superconductor into an insulating Bose glass. The associated quantum phase transition is expected to be of second order~\cite{Fisher}.
\begin{vchfigure}[htb]
  \includegraphics[width=.6\textwidth]{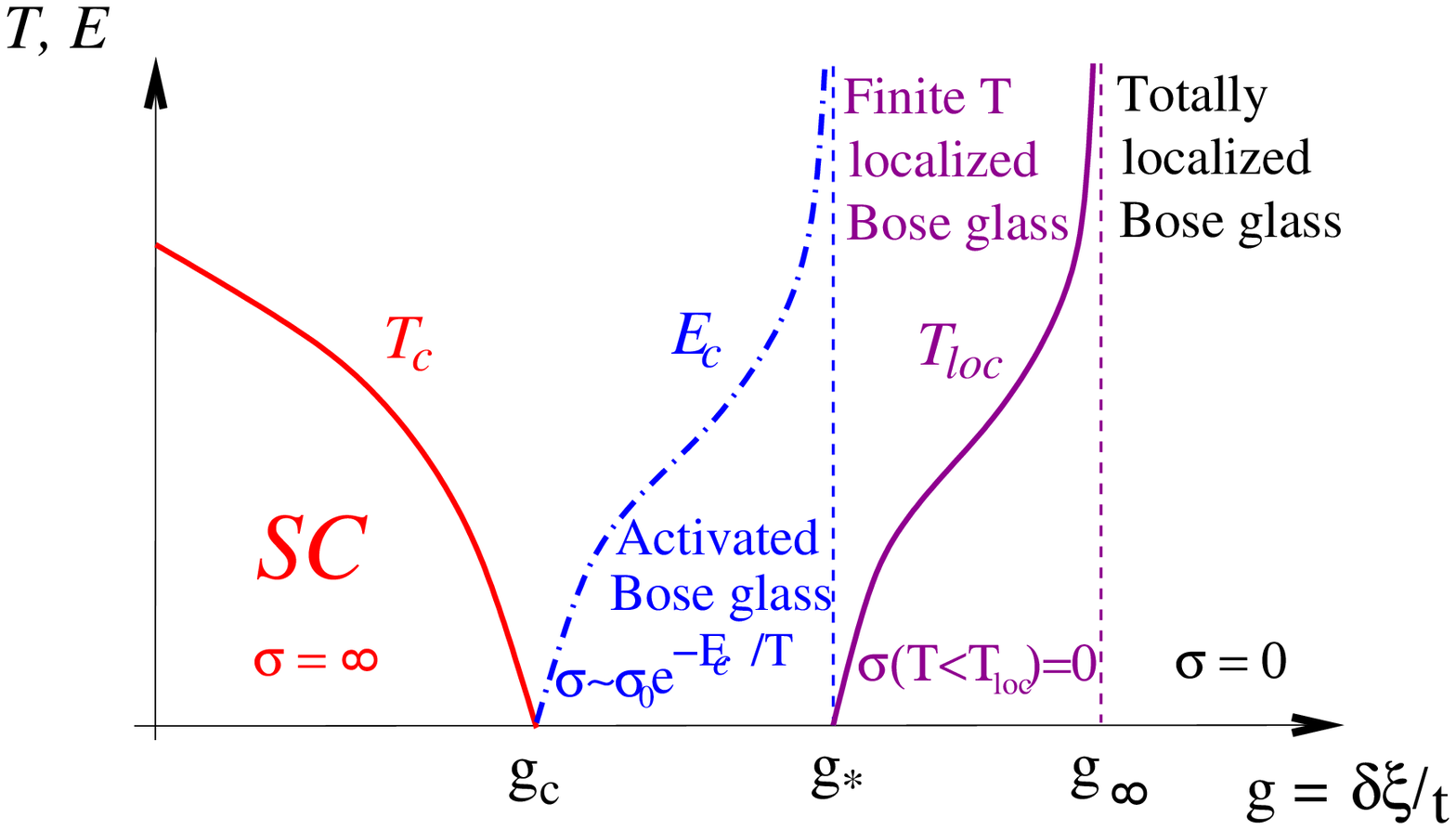}
\vchcaption{Phase diagram of the SIT and localization in the Bose glass. At strong enough disorder, $g>g_c$, superconductivity is suppressed and gives way to the insulating Bose glass. The $T=0$ mobility edge, $E_c$, rises monotonously from zero at criticality and leads to Arrhenius behavior of the conductivity at very low temperature. This part of the phase diagram should be compared to the data of \cite{activated}. At stronger disorder, $g>g_*$, non-extensive excitations localize at $T=0$, the collectively localized phase with $\sigma=0$ being stable up to a finite $T_{\rm loc}$ where a phase transition to a conducting phase occurs. At strongest disorder $g>g_\infty$ the localized phase persists up to infinite temperature.}
\label{fig:PhaseDia}
\end{vchfigure}

\subsection{Spectrum of the Bose glass}
The transport properties of an interacting insulator can best be understood by analyzing its spectrum~\cite{Berkovits, BAA, Huse}. If the excitation spectrum is discrete (point spectrum) at low energies, the system is localized and diffusion is absent. Indeed, the discreteness implies that a local excitation from an initial stationary state only couples to a finite number of eigenstates of the full system. The subsequent dynamics is thus quasiperiodic and non-diffusive, the excitation remaining local. In such a state thermalization cannot take place after the excitation, the system being in a genuine quantum glassy state~\cite{BAA}.
In contrast, a continuous spectrum reflects the fact that local excitations couple to infinitely many eigenstates of the system, so that diffusion, conduction and thermalization are possible.

It is clear that on the superconducting side of the transition the spectrum is a continuum. Indeed, the spontaneous breaking of the continuous $U(1)$ symmetry and the associated long range order ensure the presence of delocalized Goldstone modes down to the lowest energies in dimensions $d\geq 2$. Their delocalized nature implies a continuum in the spectrum of excitations at all temperatures. On the Bose glass side of the transition, the continuous spectrum cannot persist down to $\omega=0$, at least at $T=0$.
Indeed, if there were delocalized modes at arbitrarily low energies, one would expect Bose condensation into these modes, in contradiction to the absence of superfluidity. If condensation were avoided despite delocalization, one would have a Bose metal, a phase commonly believed not to exist in short range interacting bosons.~\cite{bosemetal}
Hence, if extended low energy states can be excluded, the phase beyond $g_c$ must be an insulating Bose glass with a discrete spectrum and localized excitations at energies below some threshold $E_c(g)$. On the other hand, assuming the SIT to be continuous, the mobility edge $E_c$ separating the discrete spectrum from the continuum must evolve continuously with $g$. In particular, it must vanish at criticality, $E_c(g\downarrow g_c)\to 0$.

It is important to establish whether or not $E_c$ is an extensive energy~\cite{BAA}.
We argue that close enough to the SIT the mobility edge $E_c$ is finite, while it diverges and becomes extensive at some larger disorder $g_*>g_c$. Consider the addition of an extra boson to the ground state of the Bose glass, whereby the boson is injected at an energy $\omega$ above the chemical potential $\mu$.
At small enough $\omega$, the extra boson cannot delocalize because there are no extended states at the chemical potential in the Bose glass.
However, one expects that if $\omega\gg t$, the extra boson creates an excitation which does delocalize, provided $g-g_c$ is small enough. Indeed, at $g=g_c$ we know that bosons at the chemical potential delocalize and establish long range phase coherence. This happens despite their mutual repulsion and the reduced phase space available to each boson. On the other hand, the phase space to hop and delocalize is significantly larger for a boson at higher energies $\omega\gg t$. Collisions with other bosons are indeed rare beyond the strip of energies within $\mu\pm t$ where quantum fluctuations in the ground state are significant. Therefore, those high energy states will remain extended beyond the disorder strength $g=g_c$, where only the more restricted states at $\omega=0$ start localizing.
Clearly, the excitation energy of these particular delocalized states are an upper bound for the mobility edge $E_c$, implying that $E_c$ remains non-extensive close enough to the SIT.

The above reasoning applies strictly only for $d>2$ where single bosons always delocalize in weak enough disorder, or to $d=2$ with spin-orbit scattering potentials for spinful bosons. One may worry that the situation is different in $d=2$ without spin-orbit scattering, since single particles always localize (see Ref.~\cite{Aleiner09}). However, localization in weak disorder in 2d is due to very subtle interference effects, which are likely to be destroyed by the remaining interaction effects with other bosons, even if the injection of an extra boson a high energy is considered. We expect this to preempt the scenario of $E_c$ jumping immediately to extensive values at $g_c$ in $d=2$, which would imply the simultaneous localization of {\em all} extended states at intensive energy.

If the disorder strength is increased beyond $g_*$ bosons injected at arbitrarily high energy will become localized, as well as any local excitation of non-extensive energy. This is indeed expected using arguments similar to Refs.~\cite{Mirlin,BAA}. Once the level spacing for bosons becomes significantly larger than their inelastic scattering rate from other bosons, a localization transition in Fock space is predicted.
A lower bound for $g_*$ can be obtained from the criterion that all single boson injections be localized at arbitrarily high energies $\omega\gg t$ above $\mu$. Since interaction effects are weak at these high energies, $g_*$ can be estimated by the location of complete Anderson localization of single bosons which float energetically far above the chemical potential of the ground state. Such an estimate provides a (presumably rather tight) bound for $g_*$.

\subsection{Finite temperature}
\begin{figure}[htb]
\includegraphics[width=.45\textwidth]{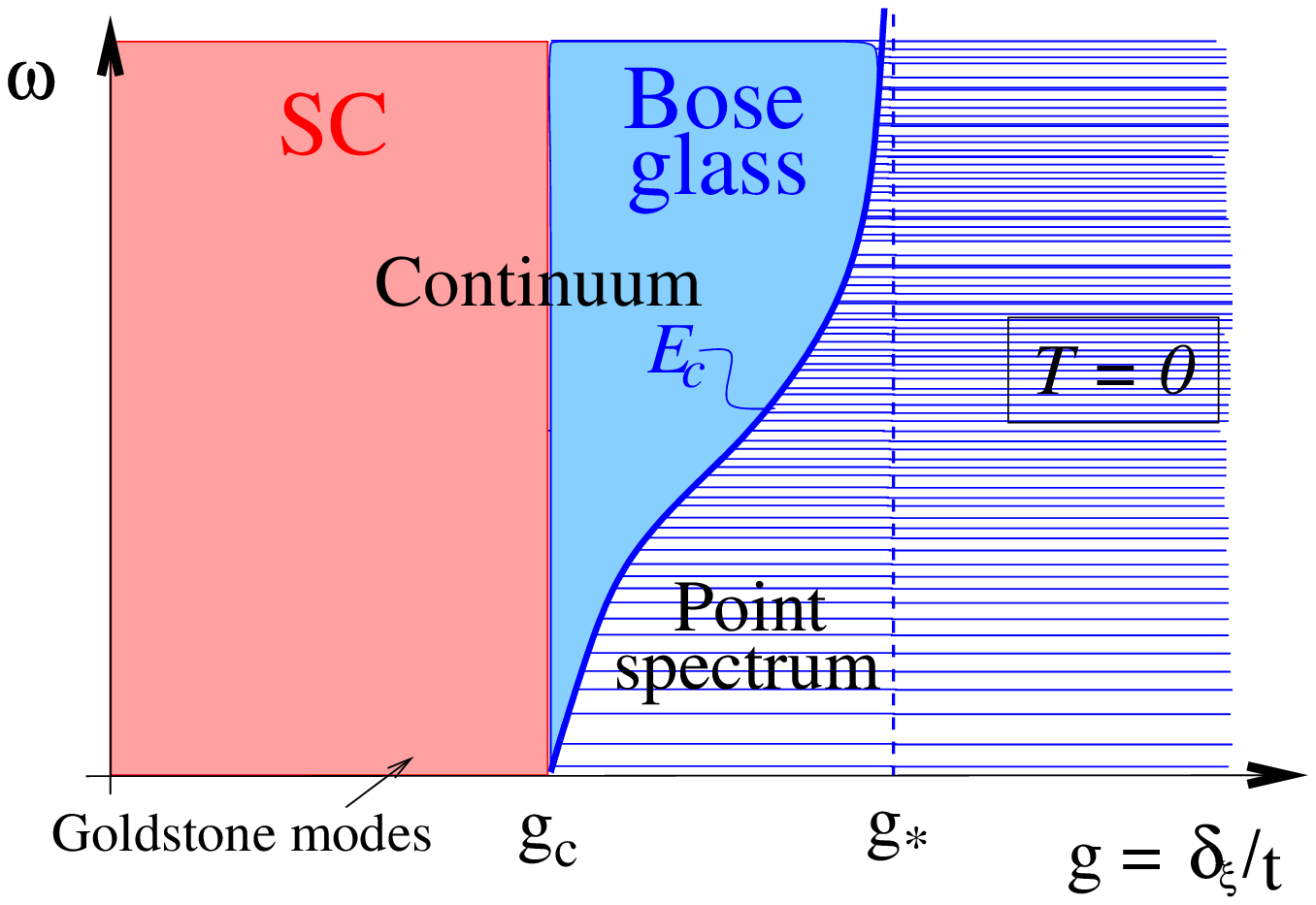}~a)
\hfil
\includegraphics[width=.45\textwidth]{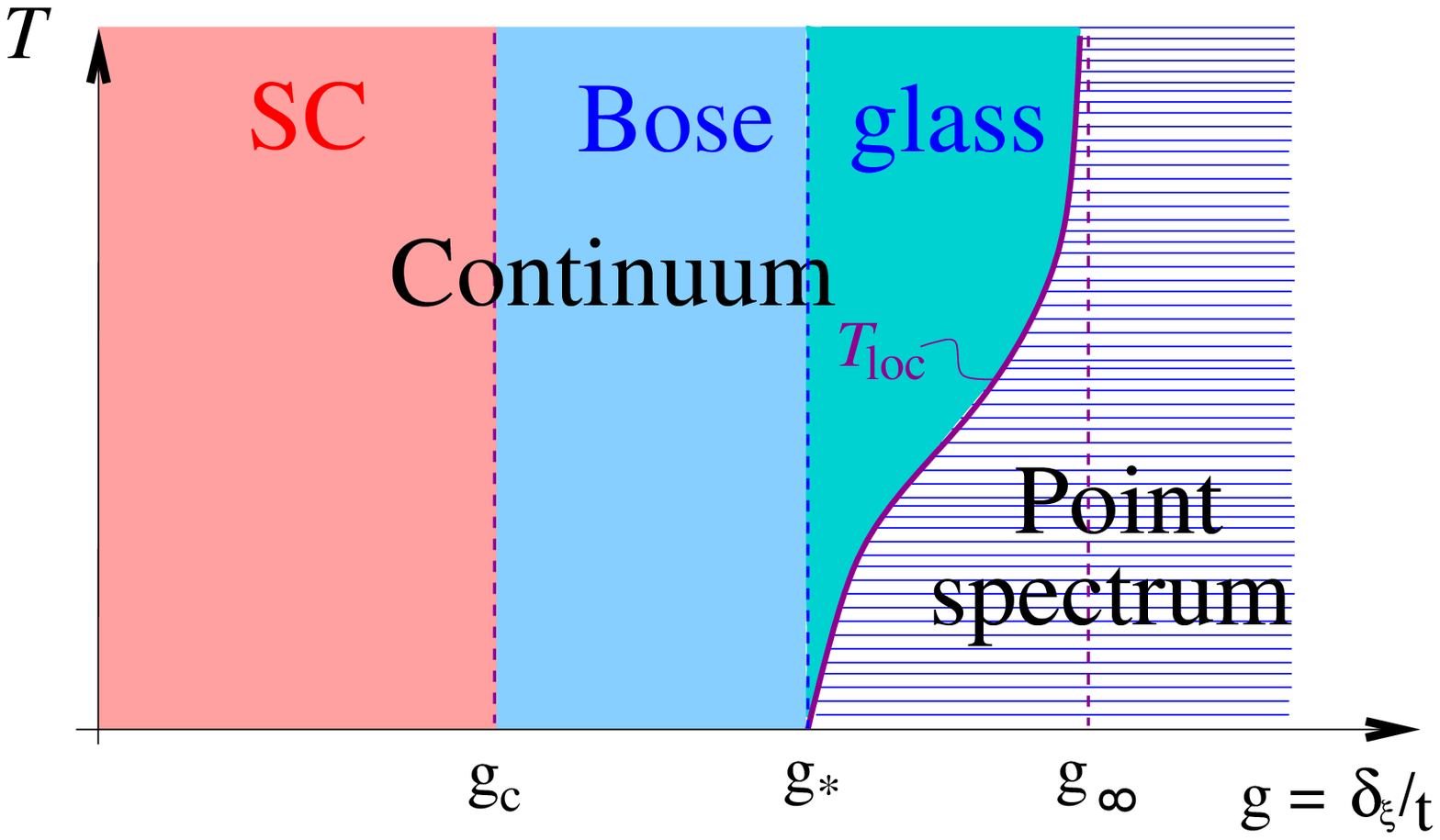}~b)
\caption{Spectrum of the Bose glass. Due to the presence of Goldstone modes the superconducting phase has a continuous spectrum at all energies and temperatures. (a) At $T=0$ the spectrum in the Bose glass is discrete at low energies, reflecting the localization of bosons. The continuum starts at an non-extensive energy $E_c$ for intermediate disorder $g_c<g<g_*$. Above $g_*$ the mobility edge is extensive. (b) At finite temperatures, the discrete spectrum only survives in the regime with an extensive mobility edge. The continuum extends down to $\omega=0$ everywhere below $g_*$ due to inelastic scattering from thermal excitations in the continuum.}
\label{fig:Spectrum}
\end{figure}

The above arguments only allowed us to conclude that the conductivity at $T=0$ vanishes in linear response, as it happens by definition in any insulator. The difference between the regimes of disorder discussed above is much more dramatic at finite temperature.
If the lower edge $E_c$ of the continuum is finite, there is a finite
density $\sim \exp(-E_c/T)$ of excitations above $E_c$ at asymptotically small temperatures. Low energy excitations which were localized at $T=0$ inelastically collide with those excitations in the continuum and thereby acquire
an inverse life time (level width). The latter is proportional to the collision probability and thus essentially to $\exp(-E_c/T)$. The whole spectrum thus becomes continuous immediately, see Fig.~\ref{fig:Spectrum}, and the conductivity will be finite for all $T>0$.
Under these conditions, the simplest conduction channel consists in the diffusion of thermally activated excitations above the mobility edge, provided the latter carry charge. Alternatively, it is also possible to have hopping transport of low energy bosons, whereby energy conservation is assured by the emission or absorption of modes above the mobility edge. At very low temperature both channels have a bottle neck arising from the exponentially small probability of finding excitations above $E_c$. Accordingly, the conductivity is expected to be asymptotically of Arrhenius form.
More precisely, if we write
\bea
\label{conductivity}
\sigma(T,g)=\sigma_0\exp[-E_A(g,T)/T]
\eea
one expects the effective activation energy to tend to $E_A(g<g_c,T\to 0)\to E_c(g)$.
Note that the above arguments assume that the mobility of excitations does not strongly decrease as $E\downarrow E_c$, and that the spectral weight in the continuum is not too inhomogeneously distributed in energy and space.
If neither assumption holds, there can be significant, but subleading, temperature corrections to the Arrhenius behavior.

Note that this conduction mechanism is entirely electronic in nature and does not depend on phonons. It thus provides a possible rationale for the occurrence of thermal run-aways in the Bose glass~\cite{overheatingtheory}. The independence from phonons should also be reflected by the pre-exponential factor $\sigma_0$. In quasi-2d samples we expect it to be a number of order $e^2/h$. This is conjectured from the analogy with the conduction at the mobility edge in (non-interacting) 2d systems undergoing a metal-insulator transition. For those systems the scaling theory of localization predicts a very weak temperature dependence of $\sigma_0 h/e^2 \approx O(1)$. Note that such a prefactor is also natural to expect if the critical conductivity is to be recovered in the limit $g\to g_c$. Empirically one indeed consistently finds prefactors of this order.

The $T$ dependence of the conductivity is entirely different for stronger disorder, $g>g_*$. Since the mobility edge is at an extensive energy, the system will remain localized at low temperature. However, as $T$ increases the number of inelastic scattering channels increases, and in some (possibly narrow) regime of disorder $g_*<g<g_\infty$ one expects diffusion to be possible above a finite localization temperature~\cite{Mirlin,BAA} $T_{\rm loc}(g)>0$. Approaching the latter from above, one expects a divergence of the effective activation energy,
$
E_A(g>g_c,T\to T_{\rm loc}(g))\to \infty.
$
Finally, if the band width of boson states is finite and interactions are short ranged, there is a maximal possible inelastic scattering rate. When it falls short of the relevant level spacing in strong enough disorder, $g>g_\infty$, the Bose glass will be localized at any temperature.

We emphasize that the non-extensive mobility edge $E_c$ is only well defined at $T=0$.
At finite $T$ inelastic collisions introduce a finite life time for excitations below $E_c$. According to the Thouless argument, diffusion is then already possible at energies $E<E_c$ where the relevant level spacing is smaller than the inverse life time. This leads to an effective lowering of the "diffusion edge" $E_A(g,T)<E_c(g,0)$ with growing temperature. Such a tendency to superactivation was actually observed in recent experiments~\cite{superactivation}.

Note that such a temperature dependence is also required to match the two regimes $g<g_*$ and $g>g_*$. At fixed temperature, $E_A$ increases monotonously with $g$, and diverges only beyond $g_*$. In particular $E_A(g_*,T>0)$ is finite, and it must be bigger than $E_A(g<g_*,T)$. This is only consistent with $E_A(g\to g_*,T=0)=\infty$, if for fixed $g$ close to $g_*$ the activation energy $E_A$ decreases with increasing temperature.


\section{Discussion}
The above scenario of the Bose glass should be experimentally relevant at temperatures where weakly phonon-assisted variable range hopping can be neglected. In order to justify the bosonic description the parameters must be such that $E_A$ is smaller than the depairing energy of electron pairs, restricting the validity of the picture to moderate magnetic fields.
The experimentally most relevant part of the phase diagram is the conducting part of the Bose glass which exhibits purely electronic activation of Arrhenius type. It would be interesting to test this scenario by confirming that transport in that regime is indeed by bosons, and by complementing transport studies with probes which are sensitive to the spectrum, such as absorption measurements or low frequency AC conductivity.

One may worry that the notion of a mobility edge $E_c$ makes an unreasonable assumption about the homogeneity of the disorder distribution tacitly implied when writing down the Hamiltonian (\ref{Ham}).
In this context it is worthwhile noticing that the typical scale of $E_c$ is of order $t$, and thus of order the bulk $T_c$ of the material. The bosonic mobility edge thus lives on a much lower scale than mobility edges in fermionic Anderson problems, which scale with the Fermi energy. Accordingly, the sensitivity with respect to inhomogeneity is much smaller at the same temperatures. Except very close to the SIT, we thus expect the bottleneck of conduction to be due to the exponentially small probability of finding excitations in the continuum, even though the spectral weight at the mobility edge of the continuum will have certain spatial variations.


In the discussion of simply activated transport channels we have assumed that the excitations at the mobility edge could carry charge. This seems plausible for the Bose glass, but is not necessarily true in more general systems close to a quantum delocalization transition. In Ref.~\cite{MullerIoffe07} a scenario for delocalized collective electronic vibrations in quantum electron glasses was considered. Those only transport energy, but no charge. If the low energy part of the continuum is entirely formed by such excitations one still expects purely electronic simple activation at the lowest temperatures~\cite{MullerIoffe07}. However, there will also be a significant range of higher temperatures where the optimal conduction channel is the variable range hopping of charged carriers, assisted by the delocalized vibrations (plasmons) of the electron glass. We suspect that a similar scenario applies to the high field regime of dirty superconducting films where most bosons depair into electrons, while the purely electronic nature of transport persists.

In the present paper we emphasized the physical picture and the ideas leading to what we believe to be a rather generic phase diagram of localization phenomena close to a transition from a disordered insulator to a conducting phase.
Calculations on the Bethe lattice, which are well controlled at high connectivity~\cite{IoffeMezard}, confirm most qualitative properties of the phase diagram.
Analytical tools allowing for a verification of the discussed scenario are presently being developed.

\begin{acknowledgement}
I would like to thank M. Feigel'man, M. P. A. Fisher, L. Glazman, D. Huse, L. B. Ioffe, V. Kravtsov, I. Lerner, M. M\'ezard and Z. Ovadyahu for useful discussions.
\end{acknowledgement}


\begin{thebibliography}{10}
\bibitem{bosemetal} However, it remains an unresolved mystery why certain supposedly bosonic systems such as Josephson junction arrays often exhibit an apparent metallic phase between the superconducting and the insulating phase.
\end{thebibliography}
\end{document}